\documentstyle[epsfig]{aa}     

\begin{document}

   \thesaurus{06         
              (08.09.2;  
               08.02.3;  
               13.25.5;  
               08.23.2;  
	       08.05.1)  
                       }
   \title{On the nature of the X-ray source 1E 
1024.0--5732/Wack~2134}

   \subtitle{}

   \author{Pablo Reig \inst{1,2,3}}


\institute{
Physics \& Astronomy Department, Southampton University, 
Southampton, SO17 1BJ, UK
\and Foundation for Research and Technology-Hellas. 711 10 
Heraklion. Crete. Greece
\and Physics Department. University of Crete. 710 03 Heraklion. Crete. 
Greece
}

\offprints{pablo@physics.uch.gr}

   \date{Received ????????; accepted ??????????}

   \maketitle

   \begin{abstract}
Two  different  models have been put forward to explain  the origin of
the   X-ray    emission    of   the    unusual    X-ray    source   1E
1024.0--5732/Wack~2134:  a high-mass  X-ray binary system  (HMXB) and a
colliding wind binary (CWB).  We present new optical and X-ray data in
an attempt to  clarify  the  nature of this  system.  The data seem to
favour  the  colliding  wind  model  since the  optical  spectra  show
characteristics  of both, a Wolf-Rayet  and an O-type  star,  implying
that these two type of stars may be present  in the  system.  The lack
of coherent  modulation  (pulsations)  and the relatively soft and low
luminosity  X-ray emission  seems to exclude the presence of a neutron
star as the body generating the X-ray  emission.  We present the first
X-ray  energy  spectrum  of this  source at  energies  above 3 keV and
discuss the implications of the spectral parameters  obtained from the
model  fitting.  We comment on the fact that the iron lines in CWB are
centred at higher  energies,  $\sim$ 6.6 keV, than those  detected  in
supergiant  HMXB, where a value of 6.4 keV is found, implying that the
degree of ionisation of iron is higher in CWB.

\end{abstract}

\section{Introduction}

The source 1E  1024.0--5732  was discovered  serendipitously  with the
{\em Einstein Observatory} by Goldwurm, Caraveo \& Bignami (1987) when
they were searching for soft X-ray  counterparts of the COS B galactic
gamma-ray  source 2CG  284-00.  Caraveo,  Bignami \&  Goldwurm  (1989)
excluded the possibility  that the X-ray emission had a coronal origin
on the  basis  of  the  ratio  between  the  X-ray  and  optical  flux
$f_x/f_{op}$   $\sim  5  \times   10^{-2}$,   which  would   imply  an
L$_x$/L$_{bol}$  much  larger  than the  average  value  of  10$^{-7}$
observed in the first X-ray studies of O-type stars (Cassinelli et al.
1981).  The source  was  visible  only in seven  Einstein  IPC  energy
channels  (0.5-3.5 keV), making any spectral  fitting  procedure  very
uncertain.  The   data   fitted   equally   well   to   a   power-law,
bremsstrahlung  and blackbody  models and therefore the X-ray spectral
parameters were affected by large  uncertainties.  Caraveo, Bignami \&
Goldwurm  (1989) also reported a periodicity  of  60.69272$\pm$0.00006
seconds,  which  they  attributed  to  the  rapid  spin  of a  compact
companion.  From  their  optical   observations  they  concluded  that
Wackerling  2134, a highly  reddened  O5 star at an  estimated  maximum
distance of 3 kpc was the most likely optical  counterpart.  Therefore
the  scenario  proposed by these  authors to explain  the nature of 1E
1024.0--5732 was that of a massive X-ray binary in which a hot O5 type
star loses mass to a spinning  neutron star.  The X-ray emission would
be due to the accretion  mechanism via the strong  stellar wind of the
optical primary.


Mereghetti  et al.  (1994)  confirmed  the  association  of the  X-ray
source 1E 1024.0-5732 with Wack~2134.  However, using data from a ROSAT
PSPC observation  they failed to detect any periodic  modulation.  The
lack of pulsations and the dependence of L$_x$/L$_{bol}$ on the poorly
constrained  interstellar  absorption  led these authors to prefer the
interpretation  of this  source  as a binary  system  consisting  of a
Wolf-Rayet  primary and an O-type  star  secondary.  In this case, the
X-ray emission would be produced as a consequence  of the collision of
their very strong stellar winds.

\section{Observations and data reduction}

\subsection{Optical spectra}

We obtained  three optical  spectra of Wack~2134 on June 20, 21 and 23,
1997  using  the 1.9m  telescope  at the  South  African  Astronomical
Observatory  (SAAO).  All  three  spectra  were  taken  with  the  ITS
spectrograph and the SITe CCD detector.  The 1200 l mm$^{-1}$  grating
No 5, blazed at 6800 \AA, was used in first  order on the  night  June
20, 1997,  whereas the 1200 l  mm$^{-1}$  grating No 4, blazed at 4600
\AA, also in first  order, was used on the other two nights.  The slit
width was set to 250  $\mu$m.  With this  set-up  the  dispersion  was
$\sim 0.5$ \AA/pixel.

Table  \ref{sp}  gives  the  journal  of  the  optical   spectroscopic
observations.  The  data  were  reduced   using  the  {\em   Starlink}
supported FIGARO package (Shortridge et al.  1997).

\begin{table*}
\begin{center}
\caption{The journal of the optical spectroscopic observations} \label{sp}
\begin{tabular}{lcccccccc}
\hline
Julian Date & Date & Wavelength range  & 
\multicolumn{3}{c}{He II $\lambda$6563/H$\alpha$}& \multicolumn{3}{c}{He II 
$\lambda$4686}\\
(2,400,000+) &	&(\AA)	&EW &FWHM  &Centre &EW &FWHM &Centre\\
\hline
50620	&20JUN97 &6180-6900	&--22.2 &45.4 &6568.3	&&&\\
50621	&21JUN97 &3970-4850	&&&	&--6.8 &20 &4690.5\\
50623	&23JUN97 &3970-4850	&&&	&--6.8 &20 &4689.7\\
\hline
\multicolumn{6}{l}{\small{EW, FWHM and centre are given in \AA.}}&&&\\
\multicolumn{6}{l}{\small FWHM not corrected for instrumental broadning.}&&&\\
\multicolumn{6}{l}{\small Errors are $\sim$ 10\%.}&&&\\
\end{tabular}
\end{center}
\end{table*}
        \begin{figure}
    \begin{center}
    \leavevmode
\epsfig{file=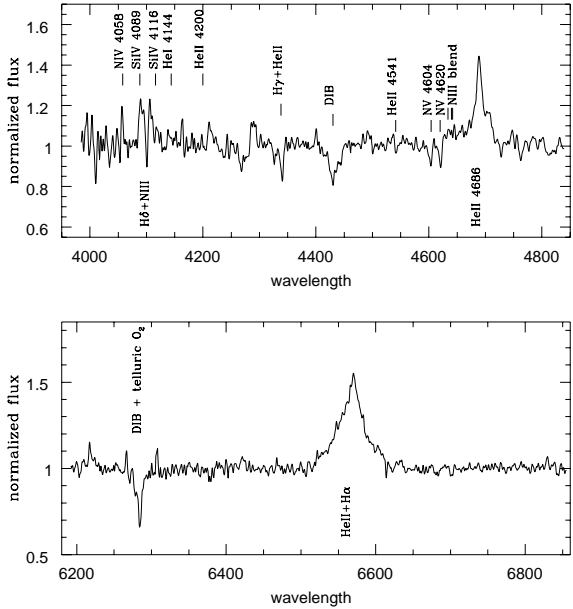, width=8.0cm, bbllx=59pt, bblly=177pt,
  bburx=558pt, bbury=700pt, clip=}
 \end{center}              
        \caption{Optical spectra of Wack~2134}
        \label{spectrum}
        \end{figure}


\subsection{X-ray data}

1E 1024.0--5732 was observed with the {\it Proportional Counter Array}
(PCA)  (Jahoda  et al.  1996)  onboard  the {\it  Rossi  X-ray  Timing
Explorer} (RXTE) on November 29, 1997 for a total on-source time of 25
ksec.  Data  reduction was carried out using the specific  package for
RXTE in {\it ftools},  whereas data analysis was done using version 10
of the XSPEC X-ray spectral  modeling  software (Arnaud 1996).  In the
analysis presented here we used the instrument  configuration known as
{\it Good Xenon}, which provides  detailed  spectra (256 channels) and
temporal ($\sim 1$ $\mu$s) information about every event that survives
background  rejection.  Good time  intervals  were defined by removing
data taken at low Earth elevation angle ($<$  10$^{\circ}$) and during
times of high particle  background.  An offset of only  0.02$^{\circ}$
between the source  position  and the  pointing of the  satellite  was
allowed, to ensure that any possible short stretch of slew data at the
beginning  and/or  end of the  observation  was  removed.  In order to
improve  the S/N ratio we used only data from the top  layer  anode of
the detectors.

Apart from 1E 1024.0--5732,  other X-ray sources may be present in the
field of view of the PCA (see Belloni \& Mereghetti  1994).  Of these,
the  emission  nebula  RCW49  is  the  most  prominent.  The  possible
contribution of this second X-ray source was taken into account in the
spectral analysis by adding an absorbed bremsstrahlung  component with
fixed parameters,  N$_H$=2 $\times$ 10$^{22}$  cm$^{-2}$ and $kT$= 0.5
keV,  as  derived  from  an   observation   with  the  {\em   Einstein
Observatory} by Goldwurm,  Caraveo \& Bignami  (1987).  This component
is not expected to affect  appreciably the spectral  parameters of the
fit  since it  contributes  only at  energies  below  $\sim$  3.5 keV,
whereas our analysis is done in the energy range 3--15 keV.


\section{Results}

\subsection{Optical spectra}

The average of the two  blue-end  spectra and the red-end  spectrum of
Wack~2134, the optical counterpart to 1E~1024.0--5732, are shown in Fig
\ref{spectrum}.  The  most  salient  features  are the  HeII/H$\alpha$
$\lambda$6563  and HeII  $\lambda$4686  lines whose equivalent  widths
(EW), full width at half maximum  (FWHM) and line centres are given in
Table~\ref{sp}.  It is clear  from  these  lines and  those of N III  (blend)
$\lambda$4634-40-42  that we are dealing  with an Of and/or a nitrogen
dominated  Wolf-Rayet  (WR) star.  The following  analysis can be done
based on the appearance and ratios of the different ions present.

\begin{itemize}

	\item The presence of hydrogen (H$\gamma$, H$\delta$) together
with  the  low  abundance  of  carbon  and  oxigen  rule  out  the  WC
classification, i.e.  a carbon  dominated  WR.  Hydrogen is  marginally
seen in late types of nitrogen  dominated  WR stars (WN) but is normal
in O-type stars (Conti \& Massey 1989).

	\item The emission seen in the He II $\lambda$4686  line and N
III  blend  $\lambda$4634-40-42  implies,  in the  case  of an  O-type
companion,  a  very   luminous   star,  that  is,  a   supergiant.  By
definition,  the O-type  luminosity  class V spectra have strong He II
$\lambda$4686  absorption.  This  combination  of He II  and N III  in
emission is denoted by an $f$  following  the  spectral  type  (Mathys
1988;  Walborn \&  Fitzpatrick  1990).  In WN stars these lines always
appear in emission.

	\item  N IV  $\lambda$4058  emission  is  stronger  than N III
$\lambda$4640,  which is a characteristic of O3-O4 spectra.  When this
situation  occurs  an $*$  following  the $f$,  that is,  $f*$ is used
(Walborn \& Fitzpatrick  1990).  WN4-5 WR stars also show strong N
IV $\lambda$4058 emission (Conti, Massey \& Vreux 1990).

	\item  If  there  is an  O-type  star,  the  absence  of  He I
$\lambda$4471 also indicates spectral type O3-O4 (Mathys 1988; Walborn
\& Fitzpatrick 1990).

	\item The lines of N V  $\lambda$4604  and  $\lambda$4620  are
strongly in absorption.  In WN stars these lines appear in emission or
very  weakly in  absorption.  Therefore  these two lines  support an O
classification.

	\item EW(He II $\lambda$4686)  $\approx-6.8$ \AA \ and FWHM(He
II  $\lambda$4686)  $\approx 20$ \AA.  The value of the EW is too high
for an O-type star but would  agree with a WN9-11  star  (Crowther  \&
Bohannan  1997), i.e, later than the  classification  implied from the
emission of N IV $\lambda$4058.  The FWHM is too high for both type of
stars but closer to a WN star than to an O-type star.

\end{itemize}

As it can be seen the spectral  classification of Wack~2134 is complex.
The nitrogen lines (absorption in N V $\lambda$4604 and $\lambda$4620;
emission in N IV  $\lambda$4058)  and the presence of hydrogen seem to
favour an O classification.  In fact, the optical spectrum of Wack~2134
is reminiscent of HD 93129A, which is classified as O3If*  (Walborn \&
Fitzpatrick 1990).  On the other hand, the very broad,  high-intensity
He II $\lambda$4686 emission would imply a WN component in the system.
One is then tempted to suggest  that both type of stars are present in
Wack~2134, forming an early-type binary.

        \begin{figure}
    \begin{center}
    \leavevmode
\epsfig{file=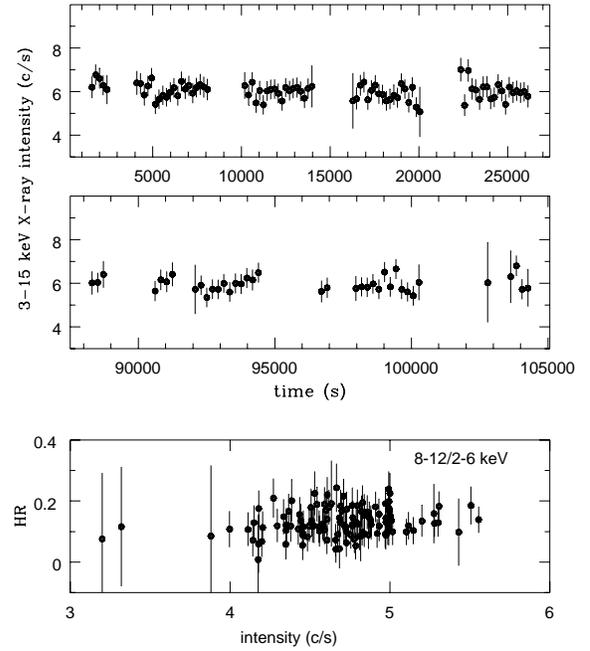, width=8.0cm, bbllx=75pt, bblly=180pt,
  bburx=560pt, bbury=715pt, clip=}
 \end{center}              
        \caption{RXTE PCA lightcurve of the entire observation. Also shown the 
HR 8-12/2-6 keV as a function of intensity. Time 0 is JD 2,450,781 at 11:26:16 
and the bin size 210 s}
        \label{pds}
        \end{figure}


\subsection{X-ray data}

Fig  \ref{pds}  shows  the  3-15  keV  PCA  lightcurve  of the  entire
observation.  The source remained fairly  constant at a mean net count
rate of 6.5 PCA counts per second.  Also shown in Fig \ref{pds} is the
hardness ratio (HR) 8-12/2-6 keV as a function of the summed intensity
in the two energy  bands.  Clearly,  the source is a soft  emitter  of
X-rays.  We searched  for the  previously  reported  60.7 ms period by
means of a Fourier analysis.  A 3500 s continuous  stretch of data was
divided into 21 intervals of 8192 bins each, with a bin size of 20 ms,
and the resulting  power averaged  together.  No evidence for coherent
or quasi-periodic  oscilations was found.  Also, a period search using
epoch folding  techniques was performed  around the 60.7 ms period but
none of the trial periods showed a high enough $\chi^2$ which could be
considered as a statistically significant detection.

In order to study the long-term X-ray  variability of 1E~1024.0--5732
we converted  all  previously  reported  intensities  into  unabsorbed
fluxes in the energy  range 3-15 keV.  The  best-fit  power-law  model
(see below), N$_H$  $\approx$  2.9 $\times$  10$^{22}$  cm$^{-2}$  and
photon   index  2.8,  was  used.  The  results   are   summarised   in
Table~\ref{int}.  For the  RXTE  observation  the  flux of the  second
source is estimated  to be about one order of  magnitude  lower in the
energy range 3-15 keV, assuming the spectral  model given in Sect 2.2.
These numbers reveal that the source shows  variability  on timescales
of years and a gradual  increase of the X-ray flux since its discovery
in 1979.

\begin{table*}
\begin{center}
\label{long_var}
\caption{Long-term X-ray variability of 1E 1024.0--5732. The intensities from 
the different instruments were converted into unabsorbed fluxes in the range 
3--15 keV by using a power-law model with $N_H$=2.9 $\times$ 10$^{22}$ cm$^{-2}$ 
and $\alpha$=2.8. The assumed distance is 3 kpc.} \label{int}
\begin{tabular}{llcccl}
\hline
Date of    &Mission \& &Intensity &Flux ($10^{-12}$ &L$_x$	  &Reference\\
observation &Instrument	& (count s$^{-1})$&erg cm$^{-2}$ s$^{-1}$)&(erg 
s$^{-1}$)  &\\
\hline
July 13, 1979	&{\it Einstein} IPC&0.036	&2.9  &3.1$\times 10^{33}$ 
&Caraveo et al. 1989\\
July 8, 1980	&{\it Einstein} IPC&0.0592	&4.8  &5.1$\times 10^{33}$ 
&Caraveo et al. 1989\\
July 29, 1992	&ROSAT PSPC	   &0.0756	&12.6 &1.3$\times 10^{34}$ 
&Mereghetti et al. 1994\\
Nov  29, 1997	&RXTE PCA	   &6.5		&15.4 &1.6$\times 10^{34}$  
&This work\\
\hline
\end{tabular}
\end{center}
\end{table*}

        \begin{figure}
    \begin{center}
    \leavevmode
\epsfig{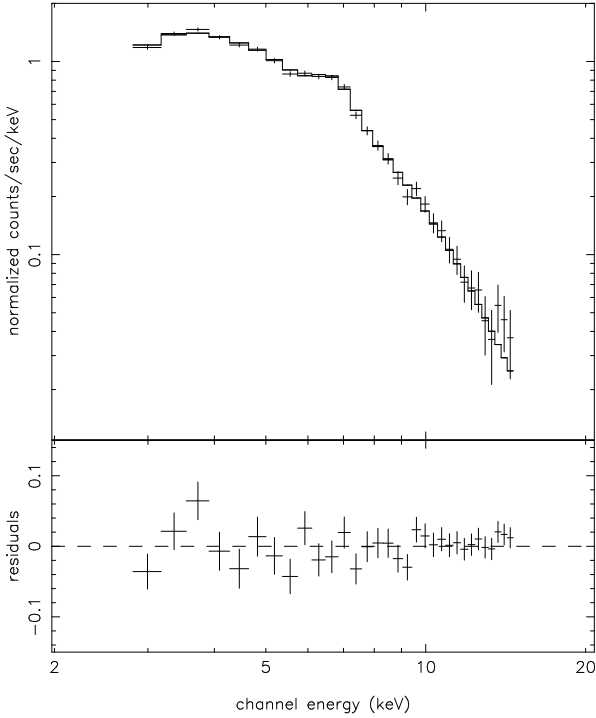}
 \end{center}              
        \caption{X-ray spectrum of 1E 1024.0--5732 in the energy range 3-15 keV. 
The straight line represents the bremsstrahlung model fit plus a Gaussian 
component, representing an iron line at 6.6 keV}
        \label{specx}
        \end{figure}

The 3-15 keV PCA RXTE  spectrum  of 1E  1024.0--5732  is shown  in Fig
\ref{specx}, where significant  emission above 5 keV can be seen.  The
spectrum is dominated  by the  emission  feature(s)  at about 6.6 keV.
Unlike X-ray  pulsars,  whose  continuum  spectral  shape is generally
represented  by a  power-law  with  an  exponential  cut-off  at  high
energies  (15--20  keV),  coupled  with low energy  absorption  and, if
necessary, an iron emission  line at $\approx  6.4$  keV, there is
not such a general model to account for the X-ray spectra in colliding
wind systems.  A two-temperature  optically thin plasma with fixed and
variable  solar  element   abundances   (MEKAL  and  VMEKAL  in  XSPEC
terminology)  was used by Skinner, Itoth \& Nagase (1998) and Corcoran
et al.  (1998) to fit the  X-ray  spectrum  of the  luminous  stars HD
50896/EZ  CMa  and  $\eta$  Carinae,  respectively.  Williams  et  al.
(1990) found that an absorbed power-law gave the best fit to the X-ray
emission of HD~193793.  The low-energy  X-ray  spectrum of the WC8+O9I
system $\gamma^2$ Vel was studied by Willis, Schild \& Stevens (1995),
who fitted a blackbody + photoelectric absorption to the observed PSPC
ROSAT data and by Stevens  et al.  (1996),  who used  combinations  of
absorbed Raymond-Smith emission models to fit its ASCA spectrum in the
energy range 0.5--5.0 keV.  Thermal  bremsstrahlung  was considered by
Williams et al.  (1997) to model the  production of X-ray photons from
the colliding-wind Wolf-Rayet system WR147.

\begin{table}
\begin{center}
\caption{Spectral fit results. Uncertainties are given at 90\% confidence for 
one parameter of interest. All fits correspond to the energy range 3--15 keV. An 
absorbed bremsstrahlung component with N$_H$=2 $\times$ 10$^{22}$ cm$^{-2}$ and 
$kT=0.5$ keV was added to all the fits (see text)}   \label{models}
\begin{tabular}{lc}
\hline
Parameter	& Value\\
\hline
\multicolumn{2}{l}{{\bf Power-law}}\\
$\alpha$				&  2.8$\pm$0.1 \\
N$_H$ (10$^{22}$ atoms cm$^{-2}$) 	&  2.9$^{2.8}_{1.1}$   \\
E$_l$(Fe) (keV)				&  6.60$\pm$0.07 \\
$\sigma$(Fe) (keV)			&  0.1 (fixed)   \\
$\chi^2_r$(dof) 			&  0.94(26)      \\
\hline
\multicolumn{2}{l}{{\bf VMEKAL}}\\
kT (keV)				& 4.8$\pm0.3$    \\
Fe ($\times$ solar)			& 0.35$\pm$0.04    \\
$\log(EM)$ (cm$^{-3}$)			& 57.28$\pm$0.09 \\
$\chi^2_r$(dof) 			& 1.68(28)       \\
\hline
\multicolumn{2}{l}{{\bf Bremsstrahlung}}\\
kT$_{brem}$ (keV)			&  5.2$\pm$0.3    \\
E$_l$(Fe) (keV)				&  6.59$\pm$0.07 \\
$\sigma$(Fe) (keV)			&  0.1 (fixed)   \\
$\chi^2_r$(dof) 			&  1.06(28)      \\
\hline
\multicolumn{2}{l}{{\bf MEKAL+ power-law}}\\
N$_H$ (10$^{22}$ atoms cm$^{-2}$) 	& $\sim$ 7   \\
kT (keV)				& 2.3$^{0.7}_{0.5}$  \\
$\alpha$				& 2.33$\pm$0.05 \\
$\chi^2_r$(dof) 			&  1.07(27)      \\
\hline
\end{tabular}
\end{center}
\end{table}

In order to fit the X-ray  continuum  we tried both,  single-component
models  (power-law,  bremsstrahlung  and VMEKAL)  and  multi-component
models  (power-law plus (V)MEKAL and  VMEKAL+VMEKAL).  The results are
summarised in Table~\ref{models}.  In the case of the single power-law
and  bremsstrahlung   models  an  additional   component   (Gaussian),
representing  an iron line at $\sim$ 6.6 keV was  necessary  to obtain
small  fit   errors.  The  width  of  the  iron  line  was   fixed  to
$\sigma$=0.1  keV since it  appeared  to be  narrower  than the energy
resolution of the PCA (18\% at 6 keV).  The line equivalent  width was
470$\pm$60  eV.  No  acceptable  fit was  obtained  using fixed  solar
element  abundances in the optically  thin plasma model.  However, the
fit improved  considerably  when we let the Fe abundance  deviate from
solar, with $\chi^2$ changing from 428 for 30 degrees of freedom (dof)
to  47  for  28  dof.  The  best-fit   temperature  was  achieved  for
$kT=4.5\pm0.2$ keV.  The X-ray luminosity in the energy range 3-15 keV
was found to be $\sim$ 1.6 $\times$ 10$^{34}$ erg s$^{-1}$, assuming a
distance of 3 kpc.

The addition of a cooler  plasma  component to the single  temperature
model with the  modified Fe  abundance  formally  improved  the fit by
decreasing  the reduced  $\chi^2$ from 1.7 to 1.5.  However, an F-test
shows that the multi-component model is not statistically significant,
with $\sim$30\%  probability that the improvement in $\chi^2$  happens
by chance.  Given the low S/N, low energy  resolution and energy range
considered   (3--15  keV)  it  is  perhaps  not   surprising  to  find
difficulties  in extracting  reliable  abundances and deriving  cooler
components.  In  contrast,  a  power-law  plus MEKAL with fixed  solar
abundances gave a good fit ($\chi^2_{\nu}$=1.07). 



\section{Discussion}

Two models are presently  competing to explain the nature of the X-ray
source  1E  1024.0--5732.  Both of them  require  the  source  to be a
binary system.  Caraveo,  Bignami \& Goldwurm  (1989)  suggested that
the system  consists of a neutron star orbiting and  early-type O star
(hereafter HMXB model).  On the other hand,  Mereghetti et al.  (1994)
proposed  an  early-type  binary  system,  in which the  primary  is a
Wolf-Rayet  star and the secondary an O-type star (CWB model).  In the
former model the X-ray  radiation is due to  accretion  of matter onto
the compact companion via a strong stellar wind, whereas in the latter
model is a consequence of colliding winds.

\subsection{Optical data}

From the point of view of theory the separation  between O-type and WR
stars does not  represent  a problem:  O stars are  assumed to be core
hydrogen  burning objects while WR have helium burning cores with very
little  hydrogen  in their  atmospheres.  This is because WR stars are
believed to be the  endpoints of evolution  of the most massive  stars.
Therefore WR stars are expected to be less massive than O-type  stars.
However,  the  distinction  between  these two type of stars  from the
point of view of the observations is not straightforward  due to their
very similar  ionising  spectrum, which is consequence of the subtypes
having  similar  luminosities  and  temperature  (Crowther \& Bohannan
1997; Crowther, Hillier \& Smith 1995).


The general criteria that have been traditionally used to separate the
two classes have been based on the full width half maximum  (FWHM) and
equivalent  width of the He II  $\lambda$4686,  with Of stars  showing
narrower  widths and lower  equivalent  widths than WN stars (Conti \&
Bohannan  1989).  However, these criteria have to be applied with care
since counter-examples can be found in the literature, such as $\zeta$
Pup, which is an O4If star  having  broader He II  $\lambda$4686  than
some late-type WN stars  (Crowther \& Bohannan  1997).  In the case of
Wack~2134  the  observed  equivalent  width  and  FWHM  of  the  He  II
$\lambda$4686  are closer to a WN star than to a single  early  O-type
star.

When other lines are considered, the spectral features  characteristic
of an O-type  star  dominate  over  those of a WN star in the  optical
spectra of Wack~2134.  Although the  classification  as a single  O3-O4
supergiant with anomalous broad, strong He II  $\lambda4686$  emission
may remain a  possibility,  it is more likely that the spectral  lines
shown in Fig~\ref{spectrum} come from both type of stars.  It is worth
noting  that,  among WR+O  binary  systems,  although  the wind of the
Wolf-Rayet companion has considerably more momentum, the O star is the
more  massive  and  likely  the  more  luminous.  Therefore  it is not
surprising to find a O-type like spectrum somehow distorted due to the
presence of the WR component.  If this  interpretation  is correct, it
would  favour the  suggestion  by  Mereghetti  et al.  (1994) that the
system is a colliding wind binary.

\begin{table*}
\begin{center}
\caption{1E 1024.0--5732/Wack~2134 compared to other colliding wind and 
supergiant X-ray binaries. {\em NS} stands for Neutron Star}  \label{comp}
\begin{tabular}{lllcccccc}
\hline
Source&Spectral	&P$_{orb}$  &distance	&Energy	range &L$_x^{max}$ &Fe line  & 
EW(Fe)&Ref.\\
name  &Type	& (days)    &(kpc)    &(keV)  &erg s$^{-1}$ &(keV) &(keV) &\\
\hline
{\bf Wack~2134}&WN5-6+O3f?&- 	&3	&3--15	 &1.3$\times 10^{34}$ &6.60 
&0.47$\pm$0.06	&1\\
HD193793 &WC7+O4-5V  &2900	&1.3	&0.5--4	 &4.0$\times 10^{34}$ &6.66 
&0.46$\pm$0.05	&2,7\\
V444 Cyg &WN5+O6     &4.2	&1.7	&0.5--4	 &7.7$\times 10^{32}$ &- &-	
&2\\
$\gamma^2$ Vel&WC8+O9I &78.5	&0.45	&0.5--10 &4.3$\times 10^{32}$ &6.9 &-	
&3\\
HD 50896 &WN5+?	     &-		&1.8	&0.5--10 &7.0$\times 10^{32}$ &- &-	
&4\\
2S 0114+650 &B1Ia+NS &11.59	&7.2	&2--10	 &3.5$\times 10^{35}$ &6.4 
&0.07--0.34	&5,8\\
4U 1700--37 &O7f+NS  &3.41	&1.8	&1--20	 &3.0$\times 10^{35}$ &6.47 
&0.07--0.40	&6\\
Vela X--1   &B0.5Ib+NS  &8.96	&2.0	&2--30	 &5.0$\times 10^{36}$ &6.42  
&0.07--0.15	&9\\
GX 301--2  &B1Ia+NS  &41.5	&1.8	&0.7--10 &1.0$\times 10^{37}$ &6.40 
&0.228$\pm$0.018 &10\\
\hline
\multicolumn{2}{l}{1: This work}&\multicolumn{2}{l}{4: Skinner et al. (1998)} 
&\multicolumn{2}{l}{7: Koyama et al. (1994)}&\multicolumn{2}{l}{10: Saraswat et 
al. (1996)}\\

\multicolumn{2}{l}{2: Pollock (1987)}&\multicolumn{2}{l}{5: Reig et al. (1996)} 
&\multicolumn{2}{l}{8: Yamauchi et al. (1990)}&\\

\multicolumn{2}{l}{3: Stevens et al. (1996)}&\multicolumn{2}{l}{6: Haberl \& Day 
(1992)} &\multicolumn{2}{l}{9: Nagase et al. (1986)}&\\

\end{tabular}
\end{center}
\end{table*}

\subsection{X-ray data}

In  Table~\ref{int}  a  long-term  increase  in the  X-ray  flux of 1E~
1024.0--5732  is  apparent.  If the system is a long period  colliding
wind binary, then colliding  wind theory  (Stevens,  Blondin \& Pollock
1992)  would  suggest  that the X-ray  luminosity  is  related  to the
separation  $D$  between  the two  components,  reaching a maximum  at
periastron, that is, $L_x \propto D^{-1}$.  Thus, a long-term increase
of the X-ray  flux may  indicate  that the  system is  moving  towards
periastron.  Such   behaviour   has  been   observed  in  HD193793,  a
P$_{orb}$=7.94  yr colliding wind system  (Williams et al.  1990).  In
support of this idea there is the fact that the flux in 1E~1024.0-5732
peaks at  $\sim$  4 keV  (Fig  \ref{specx}).  The  theory  predicts  a
hardening  of the X-ray  emission  at  periastron  due to the  greater
absorption,  with the flux maximum  moving from 1.5--2 keV at apastron
to around 5 keV at  periastron  (Stevens,  Blondin  \&  Pollock  1992).
X-rays from colliding winds are expected to show up at energies $\geq$
3--4 keV, since at lower energies the soft emission from the intrinsic
wind shocks of the individual components may dominate the spectrum.

The absence of X-ray  emission  above $\sim$ 5 keV was  considered  by
Skinner,  Itoth \& Nagase (1998) as a proof to disregard  the presence
of a compact  companion in the  Wolf-Rayet  system  HD~50896 (EZ CMa).
However, the opposite  statement,  i.e., the  presence of  significant
X-ray  emission  above 5 keV, does not  exclude  the CWB  model  since
theoretical  models and  hydrodynamic  simulations  of colliding  wind
systems predict certain level of hard (2--10 keV) X-ray emission.  The
hardest  X-rays are  expected  to come from the line of centres of the
system, where the highest  temperature  gas lies (Stevens,  Blondin \&
Pollock 1992; Willis, Schild \& Stevens 1995).  The X-ray  emission of
1E  1024.0--5732  appears  to be too soft for a HMXB.  The X-ray  flux
decreases  considerably  above 15 keV, unlike other accreting  pulsars
also observed with RXTE which shows significant emission up to 30 keV.

Likewise, the X-ray  luminosity  in the energy range 3--15 keV is also
lower, by about one order of magnitude, than the typical luminosity of
the  faintest  HMXB.  Table~\ref{comp}  shows  a  comparison  of  some
observational characteristics of 1E~1024.0--5732/Wack~2134 to other CWB
and HMXB with supergiant  primaries.  The luminosity  derived from our
RXTE data, $\sim$ 1.6 $\times$ 10$^{34}$ erg s$^{-1}$, at 3 kpc, would
make  1E~1024.0--5732 one of the more luminous colliding wind systems,
only  comparable  to HD~193793  (see  Table~\ref{comp}).  It should be
noted,  however, that the distance is poorly  constraint and the value
of 3 kpc may be  considered  as an upper  limit  (Caraveo,  Bignami \&
Goldwurm  1989).  Thus, a lower X-ray  luminosity  is possible,  which
would agree with the X-ray luminosity  normally seen in CWB (about one
order of  magnitude  lower).  Note  also  that  the  X-ray  luminosity
reported in this work  corresponds  to a broader  energy range  (3--15
keV) than normally quoted for colliding wind systems.



Single and multi-componets  models were used to fit the X-ray spectrum
of 1E  1024.0-5732  in the energy  range 3-15 keV.  The results of the
spectral fitting are summarised in Table~\ref{models}.  In the case of
the  optical  thin  plasma  emission  model  (VMEKAL)  no good fit was
achieved  by  fixing  all the  element  abundances  to  solar  values,
possibly  indicating  that the system  does not  consist of only an OB
massive  star.  By letting the Fe  abundace  be a free  parameter  the
reduced $\chi^2$  decreased to acceptable  levels ($<$ 2), although it
is still higher ($\chi^2_{\nu}=$1.7) than the other models considered.
In WR stars the outer hydrogen  envelope has been stripped away due to
the large  mass-loss  rates.  Therefore,  WR stars are  expected to be
nonsolar in their chemical composition.

The fit to the overall spectrum of CWB is generally  improved by using
two temperature  components.  A cooler  ($<$  0.7  keV)  and a  hotter
($\ge$ 2 keV) components were proven to give good fits (Skinner et al.
1998; Corcoran et al.  1998).  We tried the same approach here but the
VMEKAL+VMEKAL  model did not turn out to be statistically  significant
when compared to the nonsolar single temperature  model.  It is likely
that  this  lack  of  significance  is due to the  conditions  of  our
observation  rather than to the model  itself.  The  impossibility  of
RXTE to detect  photons below 2.5 keV makes it difficult to constrain
the temperature of a possible cooler  component and the  photoelectric
absorbing  columns.  Likewise  the low S/N and low  energy  resolution
prevent  us from  extracting  element  abundances.  Nevertheless,  the
best-fit   temperatures  of  the  bremsstrahlung  and  VMEKAL  models,
$kT\approx$  5 keV,  lie in  between  the  temperature  of the  hotter
components in other CWB -- 1.4 keV for $\gamma^2$  Velorum (Stevens et
al.  1996), 3.1 keV for  HD~50896  (Skinner  et al.  1998) and 5.9 keV
for $\eta$ Carinae (Corcoran et al.  1998).

On the other  hand, the good fit found with the  power-law  plus fixed
solar  abundances  MEKAL model  would, in  principle,  favour the HMXB
model since, on physical  grounds, we would  expect  thermal  emission
(MEKAL) from the O-type star and non-thermal emission (power-law) from
accelerated  electron in the  accretion  flow.  However, one can argue
that  the  best-fit  temperature  $kT=2.3$  keV  is  too  high  to  be
originated  in the wind of a single  O star,  for  which  much  softer
spectra  ($kT < 1$ keV) have been  observed  (Bergh\"ofer,  Schmitt \&
Cassinelli 1996).

In HMXB the presence of the iron  K$\alpha$  emission line is commonly
seen, with a variety of  strenghts.  It is  abscribed  to  fluorescent
reprocessing  by less ionized iron in  relatively  cold  circumstellar
material.  However, such a line is not exclusive of X-ray binaries but
was  reported to be present in, at least, two CWB:  HD~193793  at 6.66
keV  (Koyama  et al.  1994)  and in  $\gamma^2$  Velorum  at  6.9  keV
(Stevens et al.  1996).  In the case of 1E~1024.0--5732 the feature at
6.6 keV can be identified  as K$\alpha$  emission  from Fe  XX--XXIII.
From Table~\ref{comp} we see that iron lines in CWB seem to be centred
at slighter  higher  energies  than in HMXB,  which  implies  that the
ionisation  degree of iron in the wind surrounding the neutron star in
supergiant  HMXB is lower (Fe I--Fe XII) than in CWB ($>$ Fe XX).  The
iron line  equivalent  widths  also  tend to be  larger  in CWB.  This
difference  in the  energy  of the  iron  line  must  also  reflect  a
difference in the location and size of the region  responsible for the
line  emission.  The 6.4 keV  line  of  supergiant  HMXB  is  probably
emitted from a region  close to the neutron  star while the  $\sim$6.6
keV line of CWB would be  produced  in the  interface  where the winds
from the primary  and  secondary  collide,  that is, a more  extended,
highly  ionised  plasma,  located  farther  away  from the two  binary
components.  In  this   respect   the  6.60  keV  line   observed   in
1E~1024.0--5732  would  favour the CWB model for this system but would
deny it for $\eta$ Carinae (E(Fe)=6.44 keV).  However, the true nature
of $\eta$ Carinae is as yet uncertain (Corcoran et al.  1998).

It is interesting to note the absence of a high energy  cut-off in the
power-law model fit, so typical of accreting X-ray pulsars.  In bright
sources this cut-off  energy is seen at about  15--20 keV, but tend to
decrease to 5--10 keV as the luminosity decreases (Reynolds, Palmar \&
White  1993).  Thus, we would  expect  to see  such a  cut-off  in the
energy spectrum of 1E~1024.0-5732 if this source were an X-ray pulsar.

\section{Conclusion and future work}


The results from the optical and X-ray analysis presented in this work
seem  to  favour  the  CWB  model  for   1E~1024.0-5732/Wack~2134.  The
presence of O-type and WR features in the optical spectra, the lack of
X-ray  pulsations,  the low X-ray  luminosity,  the need for non-solar
abundances,  the  relatively  soft spectrum and the fact that the iron
line is centred at $\sim$6.6 keV are  difficult to reconcile  with the
accreting binary model.

Further  monitoring  of the X-ray flux is required to confirm  whether
the system is approaching  periastron.  Radio  observations are needed
to search for  non-thermal  radiation.  The  detection of such type of
emission would definitively confirm the nature of the system since CWB
are expected to be emitters of such type of  radiation,  whereas not a
single X-ray pulsar has been shown to emit non-thermal  radio emission
(Fender et al.  1997).  Following the scheme by Conti, Massey \& Vreux
(1990)  I- and  K-band  spectra  would  help to  refine  the  spectral
classification of the system,  specially if a WR companion is present.
Finally, high resolution X-ray spectra at low energies would allow the
detection of different metallic lines (N, Si, Fe, Mg, S) expected from
an optically  thin plasma.  In this  respect XMM may provide the final
key to unmask the true nature of this intringuing system.

\subsection*{Acknowledgments}

The author is  grateful  to M.J.  Coe, I.  Negueruela  and T.  Belloni
for their  useful  comments  and to I.  Papadakis  for his help in the
timing  analysis  of the X-ray  data.  I also  thank  the  referee  I.
Stevens for his valuable comments, which improved the final version of
this paper.  The author  acknowledges  funding via the EU Training and
Mobility Research Network Grant ERBFMRX/CP98/0195.  The data reduction
was partially carried out using the Southampton University node, which
is founded by PPARC.

\end{document}